\documentclass{elsarticle}

\usepackage[T1]{fontenc}
\usepackage{setspace}
\usepackage{graphics}
\usepackage{amssymb}
\usepackage{color}

\begin{document}

\begin{frontmatter}
\title{Structure determination of liquid carbon tetrabromide via a combination of x-ray and neutron diffraction data and reverse Monte Carlo modeling}
\author{L\'aszl\'o Temleitner}
\ead{temleitner.laszlo@wigner.mta.hu}
\address{Wigner Research Centre for Physics, Hungarian Academy of Sciences, Konkoly-Thege M. \'ut 29-33, 1121, Budapest, Hungary}

NOTICE: this is the author's version of a work that was accepted for publication in Journal of Molecular Liquids. Changes resulting from the publishing process, such as peer review, editing, corrections, structural formatting, and other quality control mechanisms may not be reflected in this document. Changes may have been made to this work since it was submitted for publication. A definitive version was subsequently published in \textit{Journal of Molecular Liquids}, \textbf{197}(2014), 204-210, DOI:10.1016/j.molliq.2014.05.011

\begin{abstract}

In order to reveal the atomic level structure of liquid carbon tetrabromide, a new synchrotron x-ray diffraction measurement, over a wide momentum transfer (Q-)range, has been performed. These x-ray data have been interpreted together with a neutron diffraction dataset, measured earlier, using the reverse Monte Carlo method. The structure is analysed on the basis of partial radial distribution functions and distance dependent orientational correlation functions. Orientational correlations behave similarly to other carbon tetrahalides. Moreover, the information content of the new x-ray diffraction data set, and in particular, of the varying Q-range, is also discussed.

Only very small differences have been found between results of calculations that apply one single experimental structure factor and the ones that use both x-ray and neutron diffraction data: the latter showed slightly more ordered carbon-carbon radial distribution function, which resulted in seemingly more ordered orientational correlations between pairs of molecules. Neither the extended Q-range, nor the application of local invariance constraints yielded significant new information. For providing a simple reference system, a hard sphere model has also been created that can describe most of the partial radial distribution functions and orientational correlations of the real system at a semi-quantitative level.

\end{abstract}

\end{frontmatter}

\section{Introduction}

The covalently bonded tetrahedral shape molecule, carbon tetrabromide (tetrabromomethane, $CBr_4$) has become a model system for studies on plastic (or orientationally disordered) crystalline phase among carbon tetrahalides. Its diffraction pattern shows only a few, not too intense Bragg-peaks; on the other hand, a significant amount of diffuse scattering contributions appear\cite{dolling_79}. To describe this behaviour various models have been refined in the past \cite{more_77, coulon_80, folmer_08, temleitner_10}. Apart from its plastic phase, $CBr_4$ has several high pressure modifications, including phase III that appears to be also plastic crystalline\cite{anderson}. Studying these exotic variations by total scattering powder diffraction may only be possible in the future, by instrumentation developed very recently for synchrotron x-ray diffraction and high-pressure techniques.

The liquid phase of $CBr_4$ is stable between 365 and 463 K\cite{crc}. In contrast to the solid phases of carbon tetrabromide, only few experiments were performed by neutron\cite{dolling_79, bako_97, temleitner_10} and x-ray\cite{folmer_08} diffraction on the liquid. Two of them \cite{dolling_79, folmer_08} (including the only published x-ray result until now) compared qualitatively the pattern of different phases; however, detailed analyses of the liquid phase was not their subject. The first theoretical investigation\cite{montague_83} using diffraction data was performed by the help of reference interaction site model\cite{rism}, which estimated the atomic radius of bromine atoms.
A more detailed analysis performed on measured diffraction datasets by Reverse Monte-Carlo (RMC) simulation was provided by Bak{\'o} et al.\cite{bako_97}, who reported all of the three partial radial distribution functions and made a simple analysis of orientational correlations. Their most important result was that they ruled out the formerly proposed Apollo-model\cite{apollo} from the most probable orientations; similarities between plastic crystalline and liquid phase were also surmised.
A more recent contribution\cite{rey_09} discussed similarities to $XCl_4$ liquids by orientational correlations based on molecular dynamics simulation and Rey's classification of mutual orientation of tetrahedral molecules\cite{rey}. This classification has been proven to be useful by numerous recent studies for liquids containing tetrahedral molecules\cite{pusztai_08,morita_09,poth_09tivcl, poth_09, caballero_12, poth_13}. Another recent examination\cite{temleitner_10} provided a systematic study of orientational correlations (again, by means of Rey's classification) in liquid and two crystalline phases by neutron diffraction, proving the similarities of orientational correlations between the plastic crystalline and liquid phases.

For the full determination of the two-particle level structure of a disordered material (containing $n$ different kinds of atoms) purely from diffraction experiments, we must be able to perform $n(n+1)/2$ independent diffraction measurements, varying the scattering parameters of each element in the material in question (wherever such a variation is feasible). However, by applying theoretical considerations, or by performing some sort of structural modeling, the required number of measurements might be reduced (see, e.g. \cite{pusztai_08,morita_09}).

Throughout the present work, effects of the new measured x-ray data, its use together with earlier neutron diffraction results\cite{temleitner_10}, and the applied momentum transfer range will be discussed, in terms of atomic radial distribution functions and orientational correlations between molecules. The feasibility of analysing the structure of liquids of tetrahedral molecules based on only one single diffraction experiment, using molecular constraints and RMC, have been discussed by many authors\cite{bako_97,jovari_01,temleitner_10}. This approach has been proven valid \cite{poth_09} for several $XCl_4$ liquids experimentally. However, such a validation is lacking for the case of $CBr_4$. Also, the latest publication on the subject\cite{temleitner_10} used an assumption based on earlier work\cite{gereben_95}, that the applied Q-range (up to only 8~\AA$^{-1}$) is sufficient for describing intermolecular correlations; this assumption also needs to be verified.

\section{Experiment}

The sample, provided by Sigma-Aldrich, contained 99\% pure $CBr_4$ polycrystals at room temperature. 

The X-ray diffraction experiment has been carried out at the BL04B2 high-energy x-ray diffraction beamline\cite{bl04b2} of the Japan Synchrotron Radiation Research Institute (JASRI/SPring-8, Hyogo, Japan). The incoming photon energy was chosen to be 37.65 keV (corresponding to a wavelength of 0.329{\AA}). The capillary transmission geometry with single HPGe detector (in the horizontal plane) setup has been used.

The powdered sample was filled into a 1 mm diameter, thin walled borosilicate glass capillary (GLAS M\"uller, Germany)  mounted in a Canberra vacuum furnace, available at the beamline. The liquid phase measurement was performed at $397.6\pm0.8$~K, recording the intensities of scattered photons by a germanium detector and the incoming beam by monitor counter. To optimize the performance of the experimental apparatus, the patterns have been recorded in four, slightly overlapping, segments, differing by the incoming beam width and height. After the measurement on the sample, scattered intensities of the empty capillary were also recorded by the same conditions. 

Raw intensities were normalized by the monitor counter, corrected for attenuation, polarization, and empty capillary intensities. Then, the whole pattern was reconstructed from the segments, scaled in electron units and corrected for Compton-scattering contributions following the standard procedure\cite{eval}.

\section{\label{sec:simulation}Reverse Monte Carlo modeling}

Series of Reverse Monte Carlo simulations (\texttt{RMC\_POT}\cite{rmcpp,locinv}) have been performed, in order to reveal the information content of the diffraction datasets beyond the already known evidences, such as molecular parameters and density (see Table \ref{tab:runs}). Taking into account solely the molecular geometry and the density, two hard sphere Monte Carlo simulations have been started: the atomic parameters of \emph{HS0} were identical to RMC calculations with experimental data (see below). In contrast, atomic radii closer to reality have been applied in the \emph{HS1} model. Of the experimentally constrained simulations, three runs were performed using both experimental (x-ray and neutron diffraction) data: \emph{NXl} and  \emph{NXlli} made use of the entire measured Q-range of both datasets, whereas in \emph{NXsh} the maximum momentum transfer values were identical for the two experimental datasets. For the 'neutron-only' case the earlier simulated configurations\cite{temleitner_10} (\emph{N}
), as well as a new simulation (\emph{Nli}) were considered.

A more detailed description of the Reverse Monte Carlo method can be found elsewhere\cite{rmc,rmcpp}, here only the way of calculation of the (total scattering) structure factor ($F(Q)$) is shown. During the RMC calculation, the partial radial distribution functions (prdf, $g_{xy}(r)$) are calculated from atomic coordinates. The prdf's are then Fourier-transformed to obtain the corresponding partial structure factors ($S_{xy}(Q)$):
\begin{equation}
 S_{xy}(Q)-1=4\pi \varrho \int_0^{\infty} r^2 \left( g_{xy}(r)-1 \right) \frac{\sin (Qr)}{Qr} d r,
\end{equation}
where $\varrho$ denotes the number density. To calculate the total scattering structure factor belonging to a given experiment, the partial structure factors are summed by proper coefficients ($w_{ij}$). Coefficients depend on concentration ($c_i$) and atomic form factors\cite{kirfel} for x-rays ($f_i(Q)$), and Q-independent scattering lengths\cite{ncoherent} for thermal neutrons as follows:
\begin{equation}
 w_{ij}=\frac{c_i c_j f_i(Q) f_j(Q)}{\left( \sum c_i f_i(Q)\right)^2} (2-\delta_{ij}),
\end{equation}
where $\delta_{ij}$ is the Kronecker symbol. The applied coefficients are shown in Table \ref{tab:weight} for both experiments.

Concerning the density of liquid $CBr_4$, earlier works performed calculations using atomic number densities of 0.031~\AA$^{-3}$ \cite{montague_83, bako_97} and 0.02688~\AA$^{-3}$ (\cite{rey_09}, referring to \cite{yaws}); these values correspond to bulk densities of 3.41 and 2.961 $g/cm^3$, respectively. Interestingly, the former value is identical to that of liquid $CCl_4$ (0.031~\AA$^{-3}$ \cite{montague_83} or 0.0319~\AA$^{-3}$ \cite{poth_09}). Since the volume of the $CBr_4$ molecule is larger than that of the $CCl_4$ molecule, the number density of liquid $CBr_4$ should be lower than that of liquid $CCl_4$, unless there are some special orientations that would enable denser packing. Since no long-range orientational correlations have been found earlier\cite{bako_97} in liquid $CBr_4$ (which might result in a more densely packed structure),  it can be concluded that some of the earlier works \cite{bako_97,montague_83} actually used a number density that is far too high. Also, a RMC trial run gave worse 
agreement (with an $R_ {wp}$ about 12\% for the neutron dataset) with unphysical features in r space, in comparison with the lower density case  (see Table \ref{tab:runs}). For these reasons, in each calculation reported here, the lower density value was taken.

For each model, the simulation box contained 6912 molecules with an atomic number density of 0.026888~\AA$^{-3}$, which provided 54.36~\AA{} half box length. To maintain the geometry of the molecules during the series of single atomic moves, fixed neighbour constraints\cite{fnc} have been applied between carbon and bromine ($1.93\pm 0.05$\AA) and between bromine and bromine atoms ($3.15 \pm 0.1$\AA) within the molecule. For pairs of atoms belonging to different molecules, closest distance constraints (cut-offs) have been used between carbon-carbon (3.5~\AA), carbon-bromine (2.5~\AA) and bromine-bromine (2.8~\AA) pairs. At \emph{HS1} run, these limits were 5.0~\AA, 3.8~\AA{} and 3.5~\AA, respectively.

The RMC procedure provides sets of configurations that agree with diffraction results within the experimental error. In order to improve the statistical accuracy of the results, series of independent particle configurations have been saved; the configurations have been taken so that they were separated by at least one successful move for each atom.

In order to favour as uniform environments of atoms as possible, the local invariance\cite{cliffe,locinv} constraint has been applied for each prdf's of \emph{Nli} and \emph{NXlli} runs up to 10.88~{\AA} in each distance bin. However, the simulation has become prohibitively slow, thus only one configuration has been saved for these two runs.

\section{Results and discussion}

\subsection{\label{ss:qgr} Results in the reciprocal space and partial radial distribution functions}

\begin{figure}[p]
\begin{center}
\rotatebox{0}{\resizebox{.8\textwidth}{!}{\includegraphics*{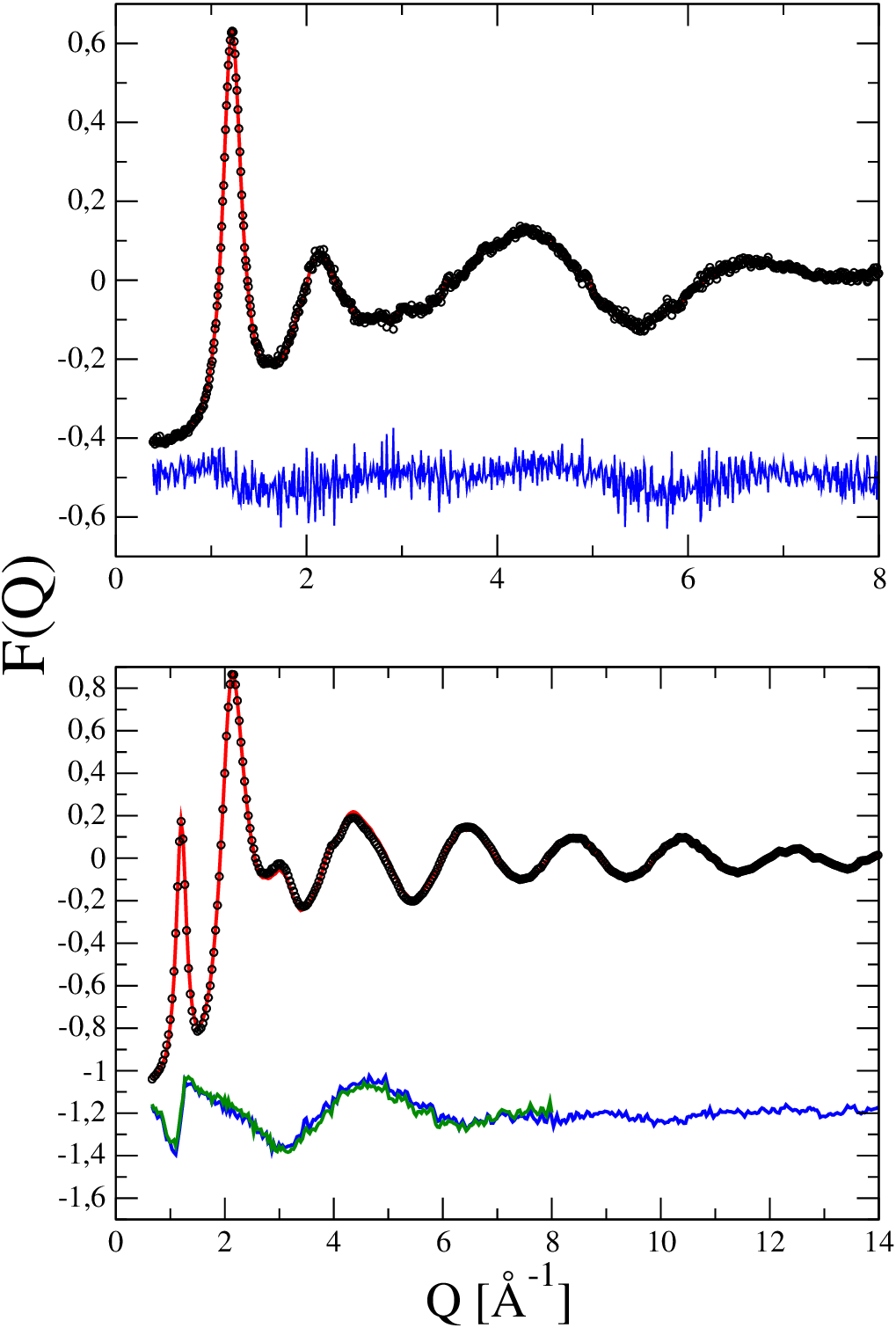}}}
\caption{\label{fig:sq} Agreement of RMC models to corrected experimental total scattering structure factors of liquid $CBr_4$. Upper panel: neutron diffraction\cite{temleitner_10}; lower panel: x-ray diffraction (this work). Circles: experimental data; red solid lines: RMC model (\emph{N} and \emph{NXl}); black solid lines: intermolecular contributions calculated from the \emph{NXl} model; blue lines: residuum calculated for the \emph{NXl} model (magnified by 5); green line: residuum for the \emph{NXsh} model (magnified by 5), respectively. The residuals are shifted along the y-axis by 0.5 in the upper, and by 1.2 in the lower panel.}
\end{center}
\end{figure}

When RMC simulations were completed, very good agreement between model and measured structure factors were obtained: the $R_{wp}$-factors are low in each case (see Table \ref{tab:runs}). The residuum (Figure \ref{fig:sq}) seems to be structureless in the case of neutron-, but somewhat more structured for the x-ray diffraction dataset. This might be the effect of small errors at the normalisation of different segments of raw x-ray diffraction data. Even though numerous runs were performed, the residuum follows similar characteristics for each of them. Thus, only the residuum of \emph{NXsh} and \emph{NXl} models are shown in Figure \ref{fig:sq}. Although there is a small difference between them, they do not differ significantly. Also, the intermolecular contributions calculated from \emph{NXl} model (figure \ref{fig:sq}) are negligible (or in the order of experimental errors) beyond 6-8 \AA$^{-1}$. These suggest that the information provided by the x-ray dataset between 8.0 and 14.0~\AA$^{-1}$ does not 
contribute much to the primary results of the simulation, i.e. to the understanding of intermolecular correlations.

\begin{figure}[p]
\begin{center}
\rotatebox{0}{\resizebox{.8\textwidth}{!}{\includegraphics*{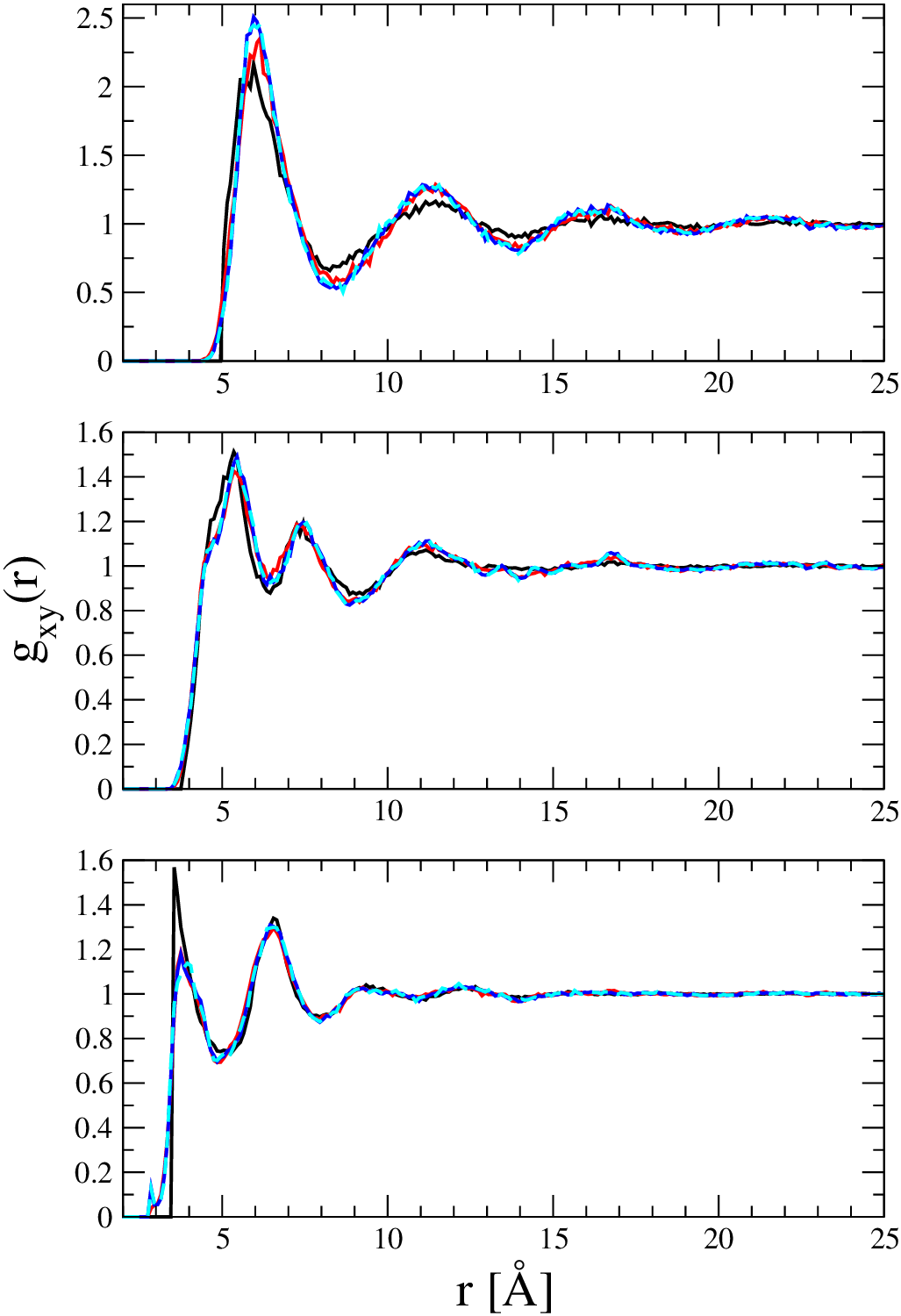}}}
\caption{\label{fig:gr} RMC refined intermolecular partial radial distribution functions of liquid $CBr_4$ as calculated from the various structural models. Upper panel: $CC$; middle panel: $CBr$; lower panel: $BrBr$. Black solid line: \emph{HS1}; red solid line: \emph{N}; blue solid line: \emph{NXsh};  cyan dashed line: \emph{NXl}.}
\end{center}
\end{figure}

Moving to the real space analysis, the intermolecular part of the RMC simulated prdf's are shown in figure \ref{fig:gr}. Having noted the only very small differences between results of simulations that apply experimental data, we can conclude, that they do not differ much from each other. Thus, the  assumption on the feasibility of structure determination based on limited Q range and on one single experimental dataset looks adequate for liquid carbon tetrabromide, too.

Going into somewhat more details, we can found slight differences between the \emph{N} and \emph{NX} runs concerning the $CC$ partial: the latter shows a more significant first maximum (around 6.0 \AA{}) and minimum (about 8.4 \AA{}). The coefficient belonging to this partial is very small for both types of data; on the other hand, there is a good contrast between x-rays and neutrons for the other two partial contributions. Note that as the $CBr$ and $BrBr$ partials are well determined (i.e., have high contributions to the total scattering structure factor), the $CC$ partial must adjust to them, via the applied molecular constraints.

A trend similar to what was found for the $CC$ prdf can be observed for the $CBr$ partial, but not for the $BrBr$ one (whose coefficient is the highest). Concerning the $BrBr$ prdf, (the simple) calculation \emph{N} reveals this function as precisely as the more sophisticated \emph{NXsh} model.
In contrast, a slight difference appears (the maximum is shifted from 3.7 to 3.9 \AA), if we compare the prdf's belonging to \emph{NXsh} and \emph{NXl} runs. That is, the applied Q-range seems to play a (very) minor role in determining prdf's of liquid carbon tetrabromide when RMC modeling is used.

The outcome of simulations using the local invariance constraint (\emph{Nli} and \emph{NXlli}) are almost identical to the corresponding unconstrained runs (\emph{N} and \emph{NXl}, respectively); these are therefore not shown in figure \ref{fig:gr}. This finding shows that the molecular geometry has been taken into account well by fixed neighbour constraints.

In order to be able to separate the structural consequences originating to the packing fraction (density and molecular shape) from those of experimental data, two hard sphere simulations have been conducted (see table \ref{tab:runs}). The \emph{HS0} results, whose intermolecular cut-off's were identical to those calculations that use experimental datasets, differ much from results based on experimental data. This suggests that effective atomic ('hard sphere') radii are larger than those applied in our RMC simulations. Also, a hard sphere model is not able to describe intermolecular prdf's in every detail. In run \emph{HS1}, the cutoff's were increased, to get more realistic results. Although agreement with results of  experimentally constrained runs are not perfect, due to arbitrary selection of intermolecular cut-offs, the $BrBr$ prdf is approximated well at distances longer than the first neighbour maximum (except around the following minimum and 
maximum distances). The applied ('hard sphere') cut-off is in agreement with an earlier RISM calculation\cite{montague_83} (where the intermolecular $BrBr$ distance was $3.5\pm 0.26$~\AA). This suggests that pairwise correlations are largely governed by the close packing of bromine atoms, in agreement with Bak\'o et. al.\cite{bako_97}. In the case of the $CC$ prdf the situation is qualitatively similar, i.e., an overall good qualitative agreement was achieved: the \emph{HS1} model is able to describe maximum and minimum positions, although estimating exact intensities is beyond the capability of the model.

\subsection{Orientational correlations}

\begin{figure}[p]
\begin{center}
\rotatebox{0}{\resizebox{.8\textwidth}{!}{\includegraphics*{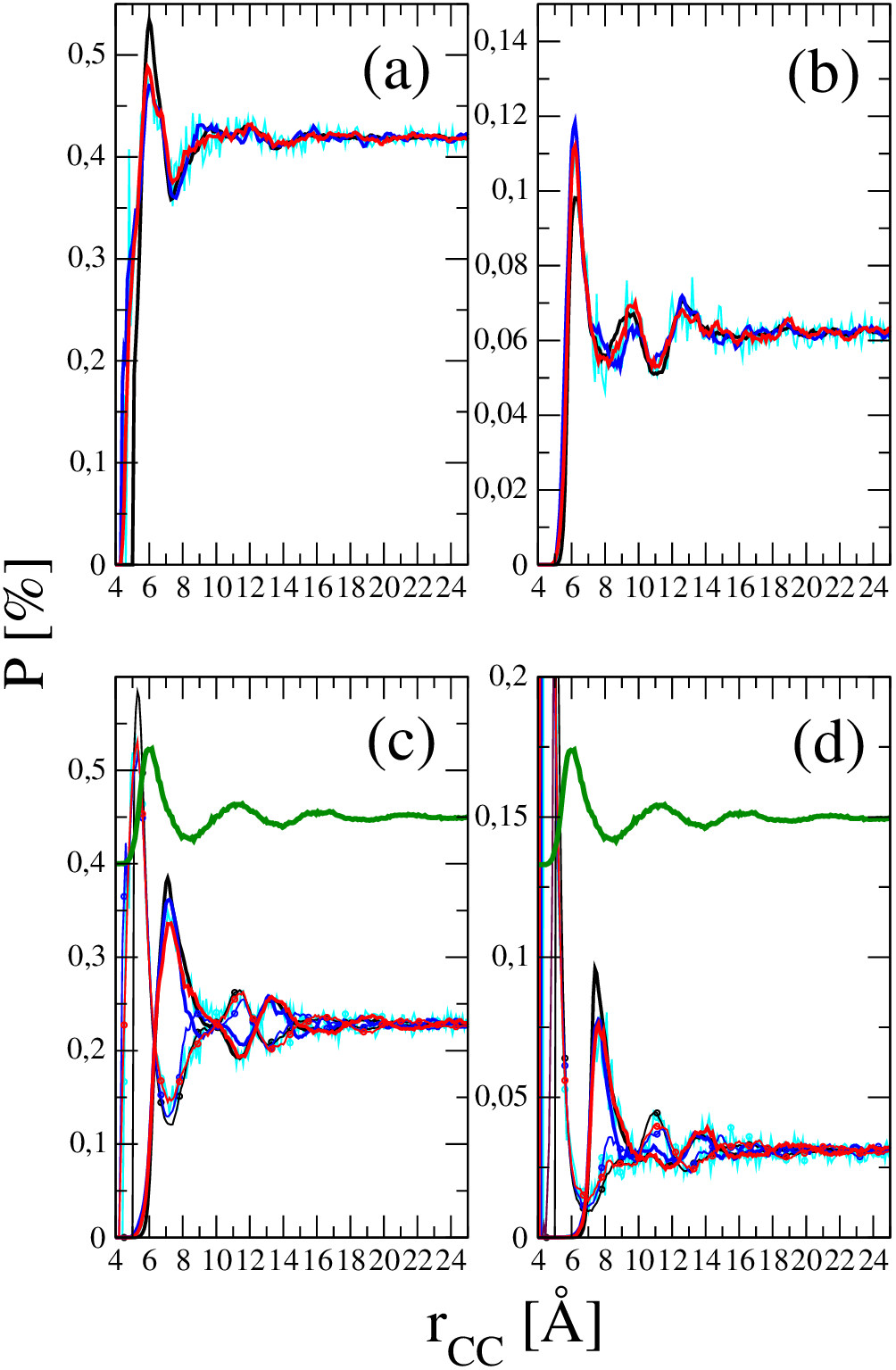}}}
\caption{\label{fig:reyn} Representation of orientational correlations by the distance-dependent probabilities of the classes of Rey, as calculated from the various model structures. (a) solid lines 2:2; (b) solid lines 1:3; (c) solid lines: 1:2, thin solid lines with circles: 2:3; (d) solid lines: 1:1, thin solid lines with circles: 3:3. Colors of different models: black lines: \emph{HS1}; red lines: \emph{N}; cyan lines: \emph{Nli}; blue lines: \emph{NXl}. Green lines: $g_{CC}(r)$ (shifted along the $y$-axis).}
\end{center}
\end{figure}

\begin{figure}[p]
\begin{center}
\rotatebox{0}{\resizebox{.8\textwidth}{!}{\includegraphics*{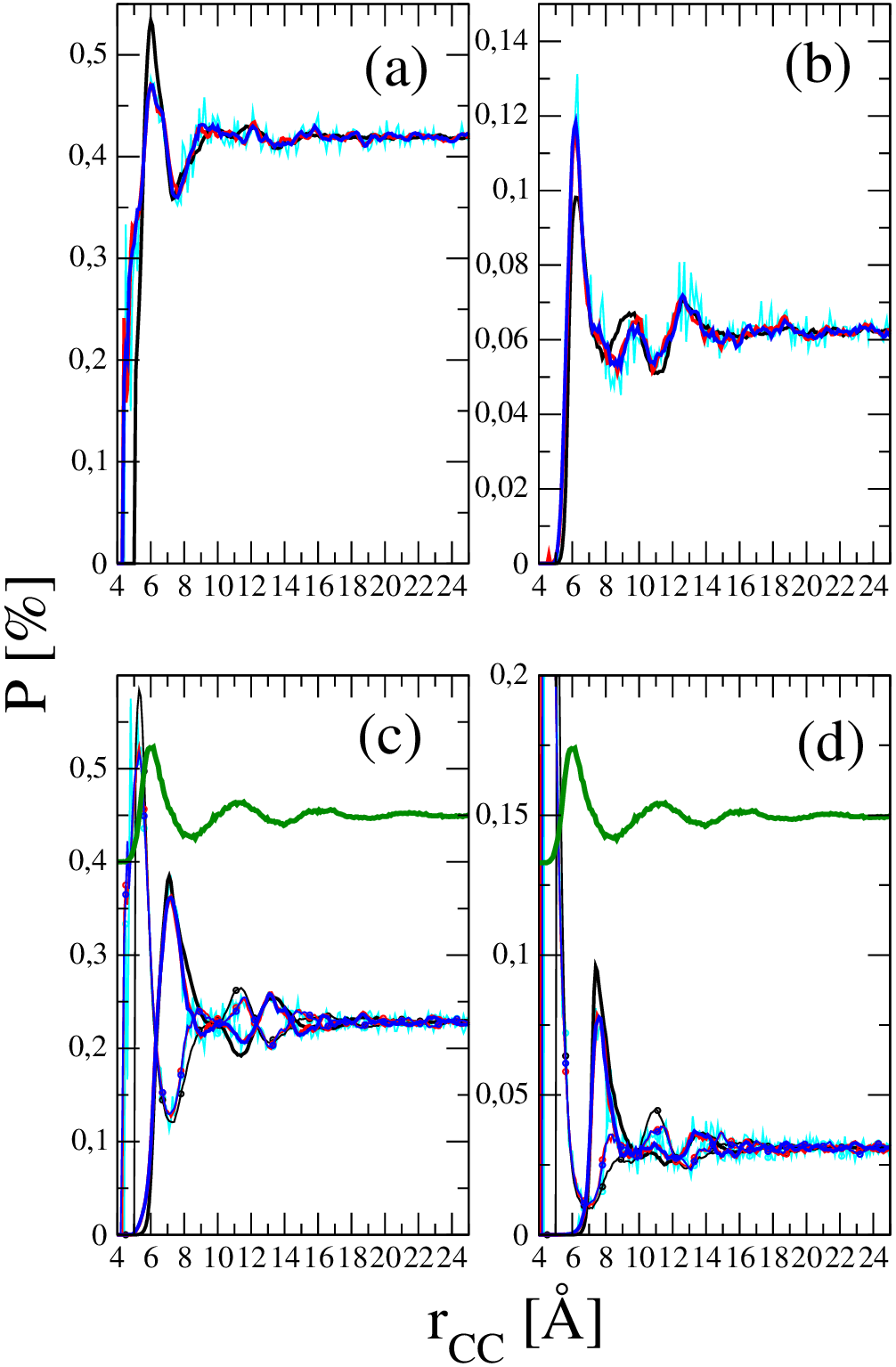}}}
\caption{\label{fig:reynx} Representation of orientational correlations by the distance-dependent probabilities of the classes of Rey, as calculated from the various model structures. (a) solid lines 2:2; (b) solid lines 1:3; (c) solid lines: 1:2, thin solid lines with circles: 2:3; (d) solid lines: 1:1, thin solid lines with circles: 3:3. Colors of different models: black lines: \emph{HS1}; red lines: \emph{NX}; cyan lines: \emph{NXlli}; blue lines: \emph{NXl}. Green lines:  $g_{CC}(r)$ (shifted along the $y$-axis).}
\end{center}
\end{figure}

In the present analysis of the mutual orientational correlations between pairs molecules, the classification of Rey\cite{rey} has been applied (figures \ref{fig:reyn} and \ref{fig:reynx}). In short, each class is described by the number of atoms belonging to each molecule between two planes containing the centres and perpendicular to the connecting line. This way, 1:1 (corner-corner), 1:2 (corner-edge), 1:3 (corner-face), 2:2 (edge-edge), 2:3(edge-face), 3:3 (face-face) classes can be formed.

By examining the resulting orientational correlations for liquid carbon tetrabromide(figures \ref{fig:reyn} and \ref{fig:reynx}), we can conclude that nearly independently of the actual model, almost all curves for a given class follow similar characteristics. The (most of the time, marginal) differences that can be found between them will be discussed below.

Generally speaking, distance-dependent probabilities of each class follow the regular characteristics shown for perfect tetrahedral\cite{rey_09, poth_09} and for $CBr_{4-x}Cl_x$ type molecules\cite{caballero_12, poth_12,poth_13} (neglecting the differences between $Cl$ and $Br$ atoms).
The first maxima for the classes follow each other from the closely contacting 3:3 pairs at the intermolecular $CC$ cutoff distance through 2:3 (5.4 \AA), 2:2 (6 \AA), 1:3 (6.2 \AA) and  2:1 (7.2 \AA) to the 1:1 (about 7.4-7.7 \AA) correlations. The most probable orientation at the asymptotic limit is the edge-edge (2:2) one, for which the correlation function becomes structureless beyond 10\AA. The corner-face (1:3) class, which would describe the 'Apollo' kind of orientational correlations, is less abundant, in agreement with earlier suggestions\cite{bako_97,rey_09}, but produce remarkable oscillations at higher distances, up to 15~\AA{}. 

In the cases of the two classes just mentioned, 4 atoms are positioned between the planes, same as the average for random orientations. For classes corner-edge (1:2) and edge-face (2:3), containing 3 and 5 atoms respectively, the deviation from the average is compensated by alternating these two orientations (see parts $(c)$ of figures \ref{fig:reyn} and \ref{fig:reynx}). The behaviour observed for liquid $CBr_4$ here agrees with the generic case\cite{rey_09} and does not show long range oscillations as found for liquid $CCl_4$\cite{poth_09}. Finally, the egde-edge (1:1) and corner-corner (3:3) type arrangements also occur as usual among tetrahalides.

Focusing at the differences between the outcome of various models, first the local invariance runs \emph{Nli} and \emph{NXlli} are examined. We can conclude that their orientational correlations do not differ from the corresponding 'standard RMC' models (\emph{N} and \emph{NXl}) (apart from the statistical noise). This reflects the fact that the 'local invariance' constraints influence primarily the \emph{intra}molecular parts of the prdf's and not the \emph{inter}molecular parts; for this reason, these (computationally expensive) constraints are not considered any further.

The next question may be whether the available/applied $Q$-range causes any difference in terms of orientational correlations. Comparing \emph{NXl} and \emph{NXsh} results in Figure \ref{fig:reynx} immediately shows that the differences are negligible. This makes sense because increasing the $Q$-range contributes mainly to fixing the molecular geometry more accurately (which is already done by fnc constraints). Thus the applicability of the original Q-range\cite{temleitner_10} for revealing orientational correlations in liquid carbon tetrabromide could be confirmed.

The additional information content of the x-ray diffraction dataset may be examined and clarified by analysing differences in the probabilities of the orientational classes belonging to \emph{N} and \emph{NXl} models (figure \ref{fig:reyn}). Several small, but distinct features may be observed. As we can expect based on the more ordered $CC$ prdf around the first maximum in the combined model, we can find differences at this distance range for 2:2, 3:3 and 1:3 orientations. As the intensities of the former ones become lower, the probability of the last one increases in comparison with the corresponding \emph{N} model. However, since the $BrBr$ prdf does not change when both diffraction data are modeled together, the more ordered $g_{CC}(r)$ influences other orientation correlations beyond the first maxima of $CC$ prdf. The most significant change is that the intensities of the first maxima of 1:2 and minima of 2:3 
classes become sharper and the probabilities get closer to the asymptotic values over the 8.5-10 \AA{} distance range. To compensate this 
behaviour, the minimum of 1:3 is shifted to 8.8 \AA{} and probabilities of 2:2 arrangements also become smaller over the 7.2-8.0 \AA{} range. This suggests that the first coordination shell is slightly more ordered than found when using only one diffraction dataset. Nevertheless, the majority of orientational correlations could be captured within errors by using one single experimental result. That is, recent RMC studies on the $CBr_{4-x}Cl_x$ family of molecular liquids\cite{poth_12,poth_13} were most certainly able to capture most structural details of the real systems.

Finally, we turn to the comparison with hard sphere results. As discussed in section \ref{ss:qgr}, only the \emph{HS1} model has been used for comparison, whose cutoffs had been set according to experimental data. As this model differs from experiment-constrained runs mainly in first coordination shell range, this kind of model is a very good approximation above 11\AA{}, where the second maxima of $g_{CC}(r)$ is found. It is important to note that the \emph{N} model is closer to \emph{HS1}, in comparison with \emph{NXl}, which contains finer details of pair correlations. In general, the \emph{HS1} model overestimates probabilities of the first maxima of 2:3, 2:2, 1:2, 1:1, while underestimates the occurrence of the 1:3 class.
The conclusion from such a high level of agreement is that orientational correlations are nearly completely determined by close packing of molecules in liquid $CBr_4$ as suggested earlier\cite{bako_97}.

\section{Conclusions}

For determining the structure of liquid carbon tetrabromide, high-energy x-ray diffraction experiments were performed, followed by an extensive series of reverse Monte Carlo calculations. Previously determined neutron diffraction data have also been made use of. Orientational correlations in the resulting structural models agree with general characteristics found for liquids of tetrahalides. During the calculations, effects of the varying Q-range and of the number and type (x-ray and/or neutron diffraction) of applied experimental data sets have been tested. It has been concluded that one single experimental data set measured over a limited $Q$-range may be sufficient for producing adequate structural models of this molecular liquid. However, using two experimental datasets yielded a better determination of the carbon-carbon prdf and gave slightly more ordered orientational correlations than found before. Increasing the upper limit of the $Q$-range from 8 to 14 \AA$^{-1}$ has not provided significant 
additional insight in terms of orientational correlations.

For the purpose of reference, hard sphere Monte Carlo runs have also been performed. The calculation that operated with realistic cut-off values was compared in detail with models based on experimental data at the level of orientational correlations. An overall semi-quantitative agreement was found, with completely matching extrema positions, while some probability differences could be detected mainly around the first maxima of each kind of arrangement. This suggested that a great deal of structural details in liquid $CBr_4$ is due to the close packing of bromine atoms, including the nearly perfect determination of the $CC$ prdf.

\section*{Acknowledgement}

This work was supported by the Hungarian Basic Research Found (OTKA) under Grant No. 083529. The synchrotron radiation experiment were performed at the BL04B2 beamline of SPring-8 with the approval of the Japan Synchrotron Radiation Research Institute (JASRI) under the Proposal No. 2010A1303. The author would like to acknowledge the Japan Society for the Promotion of Science (JSPS) for a Postdoctoral Fellowship in the period 2009-2011 and Dr. S. Kohara (JASRI) both as host researcher during the JSPS fellowship and as beamline scientist for teaching him the complete beam alignment of BL04B2. The help of L. Pusztai (Wigner RCP) is acknowledged for carefully reading the manuscript. 

\section*{References}

\bibliographystyle{elsarticle-num}
\bibliography{cbr4-liq}

\begin{table}[p]
\begin{center}
\caption{\label{tab:runs} Parameters of the Reverse Monte Carlo simulations performed. For combined runs that apply both neutron and x-ray diffraction data, the Q-ranges and R-factors are also shown.}
\begin{tabular}[c]{||c||c|c|c|c||}
\hline \hline
Run & Experimental data & Q-range [\AA$^{-1}$] & $R_{wp}$ [\%] & Saved cfg \\
\hline \hline
HS0 & none & --- & --- & 195 \\
\hline
HS1 & none & --- & --- & 184\\
\hline
N\cite{temleitner_10} & neutron & 0.39-8.0 & 4.84 & 50\\
\hline
Nli & neutron & 0.39-8.0 & 5.12 & 1\\
\hline
NXsh & neutron, x-ray & 0.39-8.0; 0.65-8.0 &  4.82; 4.89 & 68 \\
\hline
NXl & neutron, x-ray & 0.39-8.0; 0.65-14.0 & 4.84; 5.12 & 84 \\
\hline
NXlli & neutron, x-ray & 0.39-8.0; 0.65-14.0 & 5.29; 5.44 & 1 \\
\hline \hline
\end{tabular}
\end{center}
\end{table}

\begin{table}[p]
\begin{center}
\caption{\label{tab:weight} Contribution of each partial structure factor to the experimental total scattering structure factors.}
\begin{tabular}[c]{||c||c|c|c||}
\hline \hline
Experiment &  $CC$ & $CBr$ & $BrBr$ \\
\hline \hline
neutron & 3.9\% & 31.6\% & 64.5\% \\
\hline
x-ray (at Q=0) & 0.2\% & 7.09\% & 91.9\% \\
\hline \hline
\end{tabular}
\end{center}
\end{table}

\end{document}